\documentclass[a4paper,12pt]{article}
\usepackage{jheppub} % for details on the use of the package, please
\usepackage[T1]{fontenc} % if needed
\usepackage{bm,latexsym,amsmath,amssymb,amsfonts,mathtools,mathrsfs}
\usepackage{comment}
\usepackage{ulem}
\usepackage{color}

\usepackage{bm}
\usepackage{hyperref}
\usepackage{framed}
\usepackage{subfigure}
\usepackage{hyperref}   % [breaklinks,colorlinks]
\usepackage{graphicx}

\expandafter\let\csname equation*\endcsname\relax 
\expandafter\let\csname endequation*\endcsname\relax 
\usepackage{amsmath}
\usepackage{color}

\usepackage{hyperref}
\hypersetup{
    colorlinks=true,
    linkcolor=blue,
    filecolor=magenta,
    urlcolor=blue,
}
\newcommand{\T}{{\rm T}}
\newcommand{\TV}{{\rm V}}
\newcommand{\FV}{{\rm F}}

\newcommand{\be}{\begin{equation}}
\newcommand{\ee}{\end{equation}}
\newcommand{\beal}{\begin{aligned}}
\newcommand{\eeal}{\end{aligned}}
\newcommand\bea {\begin{eqnarray}}
\newcommand\eea {\end{eqnarray}}
\newcommand{\bec}{\begin{cases}}
\newcommand{\eec}{\end{cases}}

\begin{document}

\title{Hawking-Moss transition with a black hole seed}

\author[a,b,c]{Ruth Gregory}
\author[d]{Ian G. Moss}
\author[c]{Naritaka Oshita}
\author[a]{Sam Patrick}

\emailAdd{r.a.w.gregory@durham.ac.uk}
\emailAdd{ian.moss@newcastle.ac.uk}
\emailAdd{noshita@pitp.ca}
\emailAdd{sampatrick31@googlemail.com}

\affiliation[a]{Centre for Particle Theory, Department of Mathematical Sciences, 
Durham University, South Road, Durham, DH1 3LE, UK}
\affiliation[b]{Institute for Particle Physics Phenomenology, Department of Physics, 
Durham University, South Road, Durham DH1 3LE, UK}
\affiliation[c]{Perimeter Institute, 31 Caroline Street North, Waterloo, 
ON, N2L 2Y5, Canada}
\affiliation[d]{School of Mathematics, Statistics and Physics, Newcastle University, 
Newcastle Upon Tyne, NE1 7RU, UK}

\date{\today}

\abstract{
We extend the the concept of Hawking-Moss, or up-tunnelling, transitions in the 
early universe to include black hole seeds. The black hole greatly enhances the
decay amplitude, however, order to have physically consistent results, 
we need to impose a new condition (automatically satisfied for the original
Hawking-Moss instanton) that the cosmological horizon area should not increase 
during tunnelling. We motivate this conjecture physically in two ways. First,
we look at the energetics of the process, using the formalism of extended black hole 
thermodynamics; secondly, we extend the stochastic inflationary formalism to 
include primordial black holes. Both of these methods give a physical 
substantiation of our conjecture.
}

\keywords{Black holes, Instantons, Stochastic Inflation}
\preprint{DCPT-20/05}

\maketitle

\section{Introduction}

It is inevitable that quantum processes played an important role
in the very earliest stages of our universe. Possibly the most remarkable
process of all is the decay of the quantum vacuum state. This is because the change
in vacuum state can change the curvature of spacetime, and then vacuum 
decay becomes a fully non-perturbative quantum gravitational phenomenon. 
If we can provide a plausible understanding of vacuum decay in this context, 
then we may learn a little about quantum gravity.

Some time ago \cite{Hawking:1981fz}, Hawking and Moss 
noticed that the simple picture of vacuum decay 
in a system with a scalar field coupled to gravity produces strange results when
the field has a very flat potential. The usual picture of a bubble of 
true vacuum nucleating inside false vacuum with a distinct bubble wall
\cite{CDL} no longer holds: as the potential becomes flatter, the bubble wall 
becomes thicker, and the field on either side of the wall becomes closer to 
either side of the maximum of the potential barrier, until the solution 
interpolating between each side of the potential maximum can no longer
exist. Instead, it appears that the field `jumps' to the top of the potential barrier and
hence the universe undergoes a uniform jump in spacetime geometry
in which everything, up to and including the cosmological horizon, is affected. 
In a previous paper, we looked at  the way vacuum decay occurs in the
presence of a primordial black hole as the uniform field limit was approached
\cite{Gregory:2020cvy}.
In this paper, we consider vacuum decay in the
presence of a primordial black hole and a uniform scalar field.
We are led to make a new proposal, that vacuum decay is only permitted
when the cosmological horizon does not grow in size.

\begin{figure}[htb]
\centering
\includegraphics[width=0.5\linewidth]{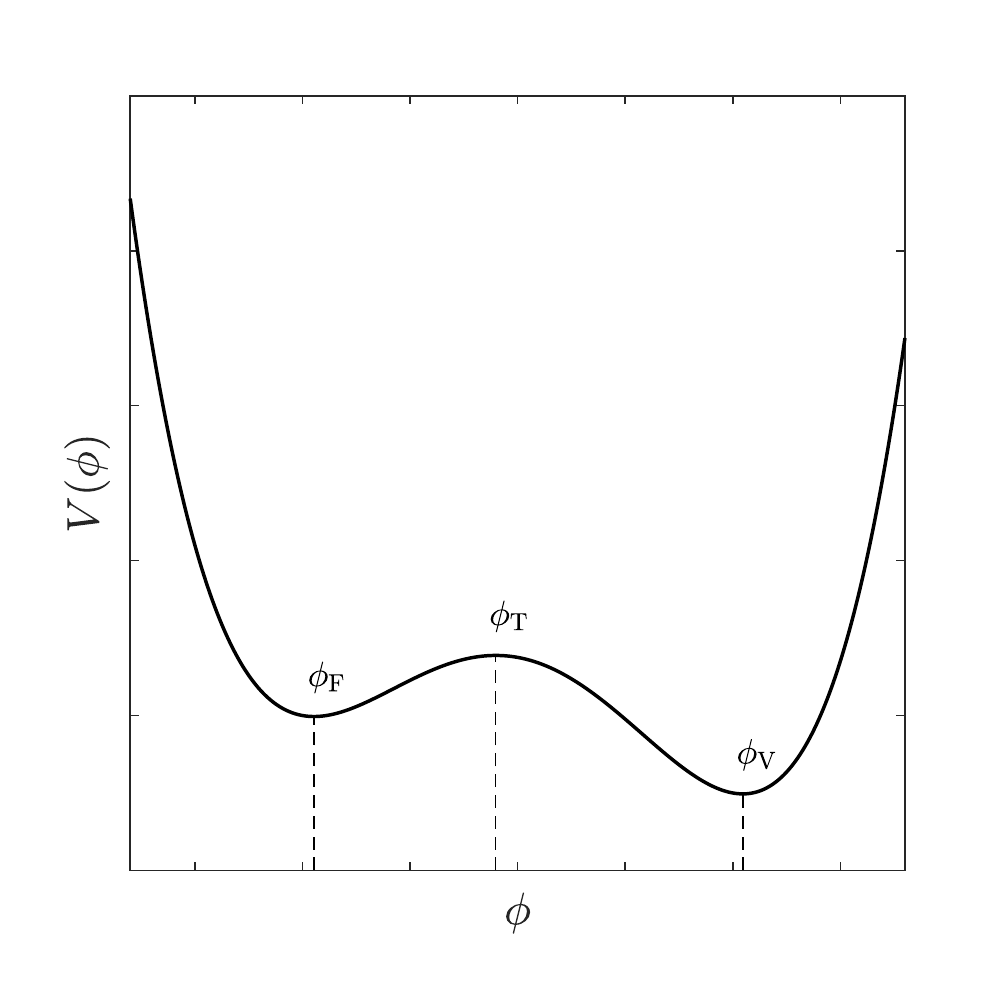}
\caption{An example potential containing a true and a false vacuum, 
located at $\phi_\TV$ and $\phi_\FV$ respectively, 
separated by a barrier peaked at $\phi_\T$.
} \label{fig:potential}
\end{figure}

The set-up is as follows: Consider a scalar field theory on a curved 
background geometry described 
by Einstein gravity, with a standard Lagrangian for the scalar field $\phi$,
\begin{equation} \label{langragian}
\mathcal{L}_\phi = -\tfrac{1}{2}\partial^\mu\phi\partial_\mu\phi - V(\phi).
\end{equation}
Since the specifics of the potential are not relevant to our discussion, let us
take a toy potential for $V$ of the form shown in Fig.~\ref{fig:potential}.
In particular, $V$ has a false vacuum located at $\phi=\phi_\FV$ and the true 
vacuum is at $\phi=\phi_\TV$.
The top of the potential barrier separating these two regions is at $\phi=\phi_\T$.
If the potential is everywhere positive, then the stable, stationary solutions 
result in a de Sitter space.

If the field is initially in the false vacuum, there is a non-zero probability to 
tunnel through the barrier to the true vacuum.
These are the bounce solutions of Coleman-de Luccia (CDL) 
\cite{coleman1977,callan1977,CDL}, which describe the nucleation of a bubble of true 
vacuum within a sea of false vacuum, i.e.\ a first order phase transition.
The bubble subsequently expands under the influence of gravitation, converting 
the false vacuum to true \cite{CDL}, at least within a
Hubble volume.

The type of transition we are interested in here occurs when the field 
undergoes a fluctuation from the false vacuum up to the top of the potential barrier.
This is known as the Hawking-Moss (HM) instanton \cite{Hawking:1981fz}, 
and involves an entire horizon volume of spacetime 
simultaneously undergoing a transition to a new state.
In situations where the CDL bounce does not exist, the only non-perturbative 
way the system can evolve into the true vacuum is via a HM bounce.

In the formal theory of vacuum decay \cite{coleman1977,callan1977}, the bounce 
solution asymptotes to the false vacuum state as the imaginary time 
becomes infinite. However, once gravity is included, {\it all} of the bounce 
solutions with positive false vacuum energy violate this condition due to the 
{\it finite} volume of Euclidean de Sitter space, but none violate this
condition more so than the HM instanton. Consequently, various attempts 
have  been made to 
understand the role of this instanton better. An early proposal was that the 
instanton solution represents the `creation of the universe from nothing' 
\cite{Vilenkin:1982de,Vilenkin:1983xq}. If this were true, then the instanton 
should play some role in the quantum wave function of the universe, and 
indeed the HM instanton gives the leading saddle-point contribution to the 
Hartle-Hawking wave function \cite{Hartle:1983ai}.

The HM instanton also plays a role in a stochastic picture of vacuum decay.
A particular feature of de Sitter space is that the large scale average of light 
fields, like the inflaton, satisfy a stochastic equation 
\cite{Starobinsky:1986fx,Starobinsky:1994bd}. 
It is therefore possible to evaluate the vacuum decay rate using 
stochastic techniques, and these reveal that the vacuum decay rate depends 
on the HM instanton in the WKB limit \cite{Linde:1993nz,Li:2007uc}. 

In yet another picture, the HM bounce can be interpreted as contributing to the
thermal ensemble of states at the Hawking temperature of de Sitter space
\cite{Brown:2007sd}. Motivated by this thermodynamical picture, it 
is important to examine the HM transition in the presence of a primordial black 
hole, which has its own additional thermodynamic profile. Indeed, it has been 
shown \cite{Gregory:2013hja,Burda:2015isa,Burda:2015yfa,Burda:2016mou} 
that the tunnelling rate for CDL bubbles is increased if a black hole is present.
Thus, a natural question to ask is how the HM instanton picture and the stochastic
formalism are altered in the presence of a black hole.

In this paper we answer this question for the case of a primordial black hole
in a single Hubble volume of the inflationary universe. 
In \S\ref{sec:HMBH} we generalise the HM instanton 
to include a black hole, and comment on how this impacts on
the instanton action. We discover that in order for the non-perturbative description
to remain well-defined we need an additional constraint on the instanton.
We therefore make the following conjecture: 

\medskip
\underline{Cosmological Area Conjecture:} 
{\it In an up-tunnelling transition, 
the cosmological horizon area can never increase}.
\medskip

Once we impose this constraint, the parameter space and
instanton actions are remarkably reminiscent of the black hole bubbles of 
\cite{Gregory:2013hja}. In the following two sections we turn to the physical 
explanation of our conjecture: In \S\ref{sec:BHTD} we
consider the thermodynamical implications of the tunnelling transition,
computing the internal energy of the false and HM states. It turns out that
the internal energy inside the cosmological horizon is directly related to the 
horizon area, thus can only increase if energy is being pumped in from 
beyond the horizon. This would correspond to an unnatural and artificially 
tuned set-up, so we conclude that an un-triggered decay cannot increase 
horizon area. In \S\ref{sec:SBH} we explore an alternate physical motivation, 
generalising the stochastic inflationary picture to include black holes. 
Using results from the analysis of slow roll inflation with black holes
\cite{Chadburn:2013mta,Gregory:2017sor,Gregory:2018ghc}, we are again 
led to the conclusion that the area of the cosmological horizon cannot increase.
We conclude in \S\ref{sec:concl}, discussing possible extensions of
our analysis.

Planck units are used throughout: $c=\hbar=k_B=G=1$.

\section{Hawking-Moss instanton with a black hole seed}
\label{sec:HMBH}

The HM instanton represents a simultaneous up-tunnelling event from a
false vacuum $\phi_F$, to the top of a potential barrier $\phi_T$. A natural
way to generalise this picture is to include a seed primordial black hole in the 
false vacuum, and to allow a remnant black hole at the top of the potential.
Typically, the masses of the black holes will be different, and this in turn will
lead to a richer set of possibilities for the tunnelling process. 

Consider a HM tunnelling event from the false vacuum at 
$\phi_\FV$ up to the top of the potential barrier $\phi_\T$, where the initial 
and final configurations contain a black hole. Assuming positive 
vacuum energy density $V(\phi)\neq 0$ for both states, the initial and final 
configurations are described by the Schwarzschild-de Sitter (SdS) solution,
\begin{equation}
ds^2 = -f dt^2 + f^{-1} dr^2 + r^2 d\Omega^2, \qquad 
f = 1-\frac{2m}{r} - \frac{r^2}{\ell^2},
\end{equation}
where the radius of curvature $\ell$ is given by,
\begin{equation} \label{curv}
\ell= \sqrt{\frac{3}{8\pi V(\phi)}}.
\end{equation}
Note that since we are interested in up-tunnelling, we always have $\ell_\FV>\ell_\T$.
In these coordinates, the range for $r$ goes from the black hole horizon $r_h$ to the 
cosmological horizon $r_c$, i.e.\ $r\in[r_h,r_c]$, where $f(r_{c,h})=0$, and the roots
can be expressed as:
\begin{equation} \label{horizons}
r_c = \frac{2}{\sqrt{3}}\ell\cos\left(\tfrac{\pi}{3}-b\right), \quad 
r_h = \frac{2}{\sqrt{3}}\ell\cos\left(\tfrac{\pi}{3}+b\right), \quad b 
= \frac{1}{3}\cos^{-1}\left(\frac{3\sqrt{3}m}{\ell}\right).
\end{equation}
The two horizons coincide at the Nariai mass $m_N$,
\begin{equation}
m_N = \frac{\ell}{3\sqrt{3}},
\end{equation}
which places an upper bound on the mass parameter, $m\in[0,m_N]$.

The tunnelling rate from the false vacuum to the top of the potential has the 
form $\Gamma \approx A e^{-B}$, where we focus on the tunnelling exponent 
$B$ rather than the pre-factor $A$. We follow Coleman and de Luccia in 
assuming that the tunnelling exponent is related to the change in 
Euclidean action $I$,
\begin{equation}
B = I_\T - I_\FV.
\end{equation}
As we stated in the introduction, there is some evidence in support of this result
from quantum cosmology and from stochastic inflation.
In SdS, the action is totally determined by the areas of the horizons 
${\cal A}_h$ and ${\cal A}_c$ \cite{Gregory:2013hja},
\begin{equation}
I = -\frac{1}{4}\left(\mathcal{A}_c + \mathcal{A}_h\right).
\end{equation}
Since each horizon is associated with an entropy ${\cal S}={\cal A}/4$, 
the tunnelling rate is related to the change in total entropy $\Delta{\cal S}$ 
by the Boltzmann formula $\Gamma=Ae^{\Delta{\cal S}}$. This links the 
tunnelling process to gravitational thermodynamics, and provides further 
support for the validity of the tunnelling formula.

The area of an horizon is $\mathcal{A} = 4\pi r^2$, so that using 
\eqref{horizons} the tunnelling exponent is,
\begin{equation}
B = \pi\left[\tfrac{4}{3}\left(\ell_\FV^2 - \ell_\T^2\right) 
- \tfrac{2}{3}\ell_\FV^2\cos(2b_\FV) + \tfrac{2}{3}\ell_\T^2\cos(2b_\T)\right].
\end{equation}
Since $\ell_\FV$ and $\ell_\T$ are fixed by the form of the potential $V$, 
we can consider the tunnelling exponent as a function of the seed and 
remnant masses, $B=B(m_\FV,m_\T)$.
The  tunnelling exponent at the extremes of the mass ranges are,
\be
\beal
B_\mathrm{HM} \equiv B(0,0) = & \ \pi\left(\ell_\FV^2-\ell_\T^2\right), \\
B(0,m_{N\T}) = & \ \pi\left(\ell_\FV^2-\tfrac{2}{3}\ell_\T^2\right), \\
B(m_{N\FV},0) = & \ \pi\left(\tfrac23 \ell_\FV^2-\ell_\T^2\right),\\
B(m_{N\FV},m_{N\T}) = & \ \pi\left(\tfrac{2}{3}\ell_\FV^2-\tfrac{2}{3}\ell_\T^2\right),
\eeal
\label{eq:Blimits}
\ee
where the HM bounce is recovered for vanishing seed and 
remnant masses, and we note that the Nariai limits for the false vacuum
and potential top are distinct, since $\ell_{\FV} \neq \ell_{\T}$.

For a black hole seed of a given mass, the remnant mass can lie anywhere in the
range $[0,m_{N\T}]$. The most probable tunnelling event will therefore be the
one with the smallest value of $B$. However, from \eqref{eq:Blimits} we see 
that if $\sqrt{\frac23} \ell_\FV < \ell_\T(<\ell_{\FV})$, it is possible for
$B(m_{\FV},0)$ to become negative for masses close to the Nariai limit. 
Negativity of an instanton action (or indeed the action dropping below one in
Planck units) indicates a breakdown of the semi-classical description 
underlying the calculation. We therefore need an additional constraint
on the tunnelling process that prevents this catastrophe.

Our conjecture (that we motivate in the subsequent sections) is to impose
that the area of the cosmological horizon should never increase
during a transition,
\begin{equation} \label{eqA}
\Delta\mathcal{A}_c = \mathcal{A}_{c\T}-\mathcal{A}_{c\FV}\leq 0.
\end{equation}
This is consistent with the idea that the instanton represents a thermal
fluctuation, because the condition implies that the fluctuation
can be contained entirely inside the original cosmological horizon.
A fluctuation that was larger than the event horizon could not
arise in a causal process.
Once we impose this constraint, we find that there is a natural cut-off 
in parameter space that keeps the instanton solutions in the range
consistent with the semi-classical approximation.

\begin{figure} [htb]
\centering
\includegraphics[width=\linewidth]{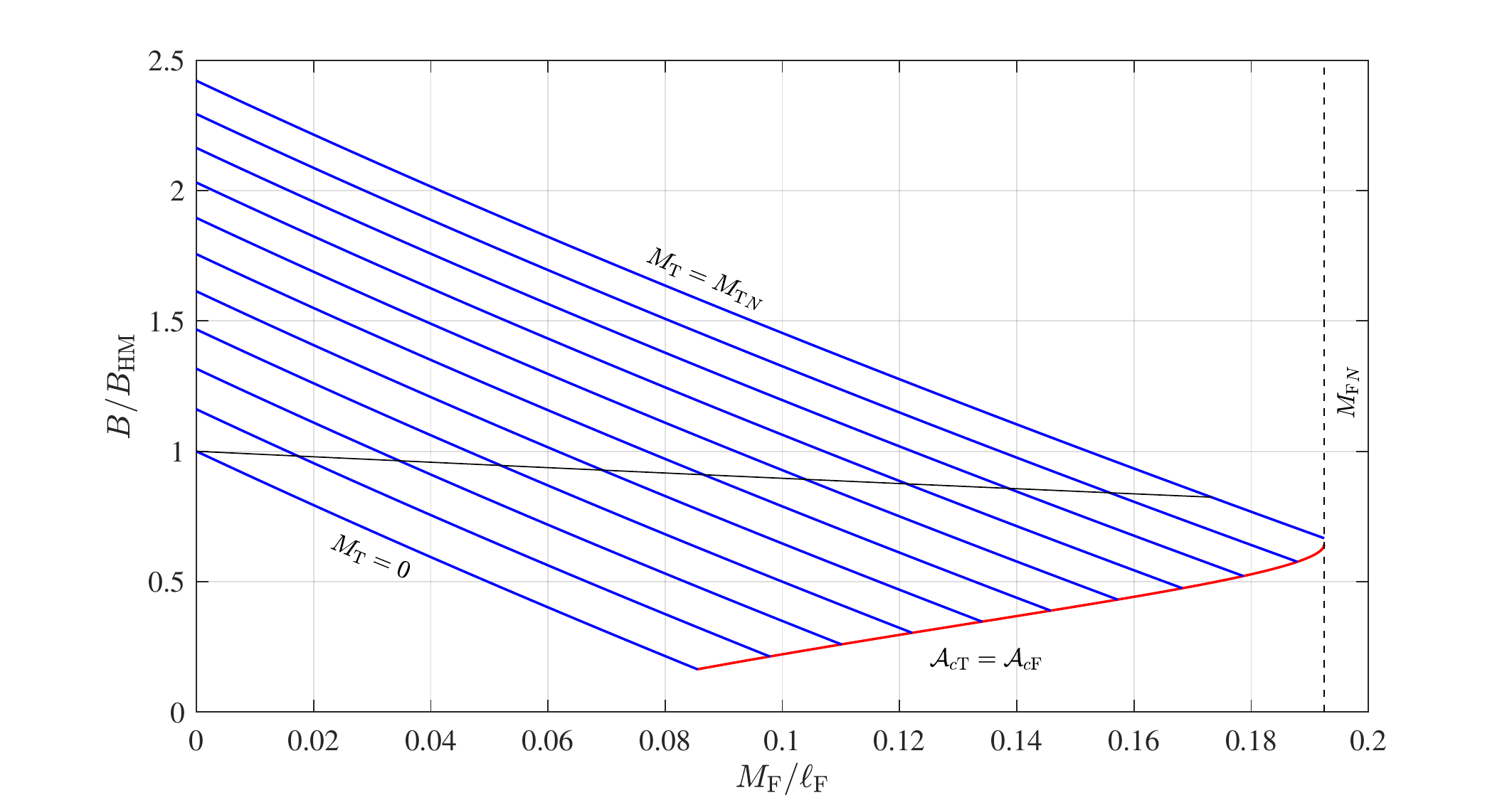}
\caption{Dependence of the tunnelling exponent $B$ on the seed mass 
$m_\FV$ for $l_\T/\ell_\FV=0.9$. The blue lines correspond to different 
values of the remnant mass $m_\T$, starting at zero and increasing in 
steps of $0.1 m_{N\T}$ up to the Nariai limit. The area of the cosmological 
horizon is conserved along the red curve and decreases above it.
Above the broken black line, the remnant black hole has a larger mass 
parameter than the seed.
} \label{fig:paramspaceMB}
\end{figure}
The parameter space of instantons is illustrated in figure \ref{fig:paramspaceMB},
where the ratio of the {\it Black-Hole-Hawking-Moss} (BHHM) instanton
action to the pure HM action is plotted as a function of the seed primordial
black hole mass, $m_\FV$. As expected, for each seed mass there is a range of
remnant masses with the action increasing as the remnant mass increases.
The blue curves in the plot show how the action varies with seed mass,
$m_\FV$, for a given remnant mass, $m_\T$. As the seed mass increases,
the action decreases until we reach the red curve boundary. This is the 
equal area curve, where the area of the cosmological horizon is the
same for initial and final states. Note that in \cite{Gregory:2020cvy},
we had imposed this as a constraint on the BHHM instantons for 
convenience.
Above the red curve, all instantons have $\Delta\mathcal{A}_c<0$,
hence are allowed, but have higher action than the equal area curve, so
are suppressed. Below the red curve, the cosmological horizon area would 
increase, which we argue is unphysical.
Thus, the condition $\Delta\mathcal{A}_c\leq 0$ provides a lower bound on the 
allowed region of the parameter space, and is pleasingly familiar from the 
black hole bubbles of \cite{Gregory:2013hja}.
The remaining bounds in the plot are fixed by 
recalling that the allowed masses in each of the SdS spacetimes are 
bounded by the appropriate Nariai mass, $m\in[0,m_N]$.
The maximal tunnelling rate occurs at the point  where $B$ is minimal, 
$m_\FV=m_C$, where the cosmological horizon areas are identical and 
the remnant mass is zero:
\be
m_C = \frac{\ell_{\FV}(\ell_{\FV}^2 - \ell_{\T}^2)}{2\ell_{\FV}^2}\;.
\label{mcrit}
\ee

Simple analytic formulae are available in the small barrier approximation 
$\ell_\FV\approx l_\T$, This approximation is equivalent to asserting that the 
height of the barrier relative to the false vacuum be small compared to its 
absolute value, i.e.\ $V(\phi_\T)-V(\phi_\FV)\ll V(\phi_\T)$.
In this case, the maximal rate is obtained for the critical seed mass
\eqref{mcrit},
%which is approximately the difference in the dS lengthscales:
$m_{C} \approx \ell_\FV - \ell_\T$.
The black hole horizon reduces approximately to the Schwarzschild value 
$r_{h\FV} \approx 2m_C$, and the value of $B$ is given by,
\begin{equation}
B_C \approx 4\pi m_C^2 \quad \Rightarrow \quad 
\left(\frac{B}{B_\mathrm{HM}}\right)_C 
\approx \frac{2(\ell_\FV-\ell_\T)}{\ell_\T}.
\end{equation}
Before moving on to examine the physics of our conjecture, note that the line 
$m_\T=m_{N\T}$ in figure \ref{fig:paramspaceMB} does not close up 
with the equal area curve at 
$m_\FV=m_{N\FV}$. This is because the cosmological horizons 
in the Nariai limit are $r_{c\FV,\T}=\ell_{\FV,\T}/\sqrt{3}$, which 
clearly do not coincide for $\ell_\FV\neq \ell_\T$.

\section{Thermodynamics of the Hawking-Moss process}
\label{sec:BHTD}

In the tunnelling scenario, we have provided additional motivation for
our result on the probability of decay as Boltzmann suppression of
an entropy-lowering transition. We now seek to explore further physical
explanations for our results. Note that the BHHM transition between
black hole spacetimes with differing cosmological constants before and 
after the transition suggests that we explore the {\it extended} black hole
thermodynamical description 
\cite{Kastor:2009wy,Dolan:2012jh,Dolan:2013ft,Kubiznak:2016qmn}, 
in which the cosmological
`constant'  determines a thermodynamic pressure $P=-\Lambda/(8\pi)$.
Of course, if $P$ is to be truly dynamical, then $\Lambda$ cannot just be
a constant term in the gravitational Lagrangian, rather, as we have here, the 
vacuum energy is determined by the expectation value of a scalar field, 
thus can obviously be allowed to vary.

The thermodynamic pressure becomes a thermodynamic charge
in the First Law, 
\be
\delta M = T \delta S + {\cal V} \delta P + ...
\label{dsfirst}
\ee
with an associated potential -- the thermodynamic volume 
${\cal V}$ -- which can be computed for each of the horizons. 
This comes with the caveat that, 
although thermodynamical relationships exist for the individual horizons, 
the temperatures of the black hole and cosmological horizon 
are unequal and the total system cannot be in thermal equilibrium.
Interestingly, as pointed out in \cite{Kastor:2009wy}, the black hole mass 
parameter, $m$, that is conventionally associated with the internal energy
of the black hole in the original formulation of black hole thermodynamics, 
in this extended formulation leads to a variable $M$ that has the interpretation 
of enthalpy, $H$, due to the first law above containing a $+{\cal V} \delta P$
term, rather than the $-P\delta {\cal V}$ term associated to ``$dU$''.
Although the extended thermodynamics of black holes is more conventionally
explored in anti-de Sitter space, where the negative $\Lambda$ gives
rise to a positive pressure, the extended thermodynamics of black holes in 
{\it de Sitter} space can equally well be considered, and was explored 
in \cite{Dolan:2013ft}, (see also \cite{Gregory:2017sor,Gregory:2018ghc})
with a Smarr relation and First Law \eqref{dsfirst} being derived. 

Now let us summarise the picture for the SdS spacetime. 
Computing the thermodynamic
parameters locally at each horizon yields
\be
M = m  \;\;\;, \qquad 
S = \pi r_{h,c}^2 \;\;\;, \qquad 
T = \frac{1}{r_{h,c}} \left ( 1 - 3 \frac{r_{h,c}^2}{\ell^2} \right)  \;\;\;, \qquad 
{\cal V} = \frac{4\pi}{3} r_{h,c}^3 \;,
\ee
however, notice that this definition of the temperature yields a negative sign,
and sometimes the modulus is taken. We will retain the signs here however
for consistency of the expressions that follow.
It proves useful to repackage these expressions in a `chemical' form, 
following \cite{Dolan:2012jh,Gregory:2019dtq} that
uses only thermodynamic charges:
\be
M = \sqrt{\frac{S}{4\pi}} \left ( 1+ \frac{8PS}{3} \right) \;\;\;, \qquad 
T = \frac1{4 \sqrt{\pi S}}  \left ( 1+ 8PS \right)  \;\;\;, \qquad 
{\cal V} = \frac43 \sqrt{\frac{S^3}{\pi}} \;.
\ee

Let us now consider the internal energy bounded by the cosmological horizon;
we can think of this as the total energy in the observable de Sitter universe. 
According to the thermodynamic expressions, this is
\be
U = M - P{\cal V} = \sqrt{\frac{S}{4\pi}} \left ( 1+ \frac{8PS}{3} \right) 
- \frac{4P}3 \sqrt{\frac{S^3}{\pi}} = \sqrt{\frac{S}{4\pi}} \;,
\ee
thus, the total internal energy of the SdS spacetime is determined by
the entropy of the cosmological horizon. 

During a decay, the only
way we can imagine the internal energy of the spacetime to increase is
if there is an influx of energy from beyond the cosmological horizon. 
This would therefore not represent a spontaneous transition between vacua, 
but would be more analogous to a stimulated decay (and one would also have 
to take account of this input in any computation of a decay amplitude). However,
it is natural to imagine that energy can be dissipated beyond the horizon,
or that the decay gives an energy neutral budget. We therefore posit that the
internal energy of the spacetime must not increase in any decay, hence 
$\delta S_c \leq 0$. This provides a natural constraint on the
space of HM instantons. As we see from figure \ref{fig:paramspaceMB},
for a given seed mass, the preferred HM instanton either has no
remnant black hole, or has a remnant, but conserves the internal energy of
the observable de Sitter universe.

\section{Stochastic tunnelling in the presence of a black hole}
\label{sec:SBH}

We now explore a very different approach,
based on stochastic inflation, to support our premise that
the cosmological horizon area decreases for the HM type of tunnelling process.
We start from de Sitter space to review some of the basic premises, and
then modify the stochastic formalism to include a population of primordial
black holes.

In the stochastic inflationary formalism, the inflaton field is averaged over large
spatial scales in a spatially flat universe to produce a `coarse grained' effective 
cosmological model \cite{Starobinsky:1986fx,Starobinsky:1994bd}.  
The effective field $\phi(t)$ evolves by a stochastic equation
\begin{equation}
3H\partial_t\phi=-\partial_\phi V+\xi,\label{sde}
\end{equation}
where $\xi$ is a gaussian random function that arises from the effects 
of small-scale quantum fluctuations. The noise correlation function 
obtained from quantum field theory can be approximated 
by a local expression with diffusion coefficient $D$,
\begin{equation}
\langle \xi(t)\xi(t')\rangle=2D\delta(t-t')=\frac{9H^5}{4\pi^2}\delta(t-t').
\end{equation}
If the field is released in the false vacuum $\phi_F$, and there is a potential
barrier with the top at $\phi_\T$, then the stochastic source in Eq. (\ref{sde}) 
pushes the field across the top of the barrier. The probability to remain
inside the barrier falls, and therefore the false vacuum decays.
The decay constant $\Gamma$ is given by a general formula 
\cite{Linde:1991sk,Li:2007uc}
\begin{equation}
\Gamma=\frac{1}{2\pi\gamma}\left(V''(\phi_F)V''(\phi_\T)\right)^{1/2}
e^{-\gamma(V(\phi_\T)-V(\phi_F))/D},
\end{equation}
where $\gamma$ is the effective friction: in our case $\gamma=H$.
This decay constant is of the form $Ae^{-B}$, with
\begin{equation}
B=\frac{8\pi^2\delta V}{ 3 H^4}\;,
\end{equation}
exactly as we have for the HM instanton when $\delta V=V(\phi_\T)-V(\phi_\FV)\ll V$.
Note that the tunnelling result only holds as long as $B\gg 1$.
If the potential is very flat, the field does a random walk and reaches $\phi_\T$ on
a timescale of $\phi_\T^2/H^3$ that does not depend on the barrier height.

We have a stochastic picture of the HM transition, but our concern is how the 
stochastic decay affects the cosmological horizon area,
${\cal A}=4\pi/H^2$. According to the stochastic inflationary formalism 
\cite{Starobinsky:1986fx,Starobinsky:1994bd}, the back reaction of the field
on the metric implies that the Hubble expansion rate, $H(t)$, varies over large scales
according to the Friedmann equation,
\begin{equation}
3H^2=8\pi V,\label{friedman}\;.
\end{equation}
Starting with the field in the false vacuum, a change in potential 
$\delta V$ induces a change 
in the horizon area $\delta{\cal A}$,
\begin{equation}
\delta{\cal A}=-{8\pi{\cal A}\over 3H^2}\delta V.
\end{equation}
Therefore, stochastic evolution to the top of the potential barrier, which we have
argued corresponds to the HM instanton transition, causes a decrease in horizon area.

To extend these ideas to black holes in de Sitter space, we need a notion of the 
slow roll equation with a black hole, together with a time and radially
dependent counterpart to the
Friedmann equation. Fortunately, this problem was addressed for a
single black hole with a slowly evolving scalar field in a
sequence of papers \cite{Chadburn:2013mta,Gregory:2017sor,Gregory:2018ghc}. 
Physically, a black hole with a slowly evolving scalar field will be
very close to a SdS spacetime, therefore the solution is expressed as
a perturbation of SdS in time. This will not necessarily yield a solution for arbitrarily
long timescales, but will give a good approximate solution in the same sense
that slow roll inflation gives a good approximation to the inflationary universe.

The first step is to identify a ``time'' coordinate in the SdS spacetime that 
will asymptote cosmological time beyond the cosmological horizon. This
is done by identifying the direction in which the scalar field rolls. The challenge
is that this coordinate must be regular at each of the 
black hole and cosmological horizon radii, $r_h$ and $r_c$
respectively. This was identified in
\cite{Chadburn:2013mta,Gregory:2017sor,Gregory:2018ghc}, and 
interpolates between the local advanced Eddington time $v$ at
the black hole horizon, and retarded time $U$ at the cosmological horizon.
The time coordinate takes the form $T=t+h(r)$, (although any rescaling of
this combination by a constant factor will also work).

In \cite{Gregory:2017sor,Gregory:2018ghc}, it was shown that, provided the
standard slow roll relations \cite{Liddle:1993fq} for the potential $V$ are 
satisfied, then $\phi$ approximately solves a modified slow roll equation:
\begin{equation}
3\gamma \frac{d\phi}{ d T}=-\partial_\phi V,\label{clas}
\end{equation}
where
\begin{equation}
\gamma=\frac{r_c^2+r_h^2}{ r_c^3-r_h^3}.
\end{equation}
As pointed out in \cite{Gregory:2017sor}, $\gamma$ has the nice 
thermodynamical interpretation as being, up to a factor,
the ratio of the total entropy divided by the thermodynamic volume 
of the intra-horizon SdS system.

In order to obtain a stochastic system, we divide space into cells, and average
as in stochastic inflation, but now include one black hole in each cell. 
Small scale quantum fluctuations will cause the field to evolve stochastically, 
and we replace Eq. (\ref{clas}) with
\begin{equation}
3\gamma \frac{d\phi}{ d T}=-\partial_\phi V+\xi.
\end{equation}
By analogy with Eq.\ \eqref{sde}, we expect the noise correlation 
function to be of the form
\begin{equation}
\langle \xi(T)\xi(T')\rangle=2D\delta(T-T').
\end{equation}
However, the particular form of the noise correlation function does not 
affect the argument which follows.

We are particularly interested in how the scalar field back-reacts
on the geometry, specifically, the area of the horizons. 
In \cite{Gregory:2017sor}, the evolution of the horizons was analysed, and
to leading order it was found that
\be
\delta {\cal A}_i = -\frac{8\pi{\cal A}_i}{3 \gamma |\kappa_i|} \delta V,
\label{horizonvariation}
\ee
where $\kappa_i$ is the surface gravity of the horizon in question, explicitly:
\be
\kappa_h=\frac{(r_c-r_h)(2r_h+r_c)}{ 2r_h(r_h^2+r_c^2+r_hr_c)}\;\;\;,
\qquad
\kappa_c=\frac{(r_h-r_c)(2r_c+r_h)}{ 2r_c(r_h^2+r_c^2+r_hr_c)}.
\ee
We see therefore that \eqref{horizonvariation} implies that under stochastic 
evolution from an initial false vacuum $\phi_\FV$ to $\phi_\T>\phi_\FV$, 
with $\delta V>0$, the horizon areas {\it decrease}. 
This confirms our general proposal that the cosmological horizon shrinks during the
up-tunnelling type of vacuum decay.

An interesting corollary is that since the same qualitative behaviour of horizon
area occurs at {\it each} horizon, the black hole violates the area theorem 
during HM vacuum decay. This could not happen for a purely classical 
process, and confirms the quantum nature of vacuum decay. 
The decoherence process of quantum fluctuations leads to entropy 
production by which the generalized 
second law may be satisfied \cite{Oshita:2017hsb}. 

\section{Conclusion}
\label{sec:concl}

We have seen that the Hawking-Moss, or up-tunnelling, types of transition
extend naturally to vacuum decays seeded by black holes, as long as we impose 
a condition that the nucleation event can be contained within the original 
cosmological event horizon, i.e.\ the cosmological horizon does not increase in area.
The vacuum decay rate is always enhanced by the black hole seed.
The mass of the remnant black hole after vacuum decay can be zero,
and the cosmological horizon shrinks, or the remnant mass is non-zero
and the cosmological horizon stays the same size. Which of these
outcomes occurs depends on the value of the seed black hole mass.

The theory of stochastic inflation was extended to include
primordial black holes in \S \ref{sec:SBH}. In stochastic inflation, the
Hawking-Moss instanton gives the leading order approximation for calculating the 
probability flux across the potential barrier.  The new theory implies 
that both the black hole and cosmological horizon areas decrease 
during up-tunnelling events. On the other hand, the stochastic picture
allows less freedom in the choice of remnant mass than does the
vacuum tunnelling picture.

It would be interesting to generalise these results to rotating black holes, 
as has been explored for black hole bubbles in \cite{Oshita:2019jan}.
It might seem that the outcomes would be similar, however there are several
important technical differences. Positivity of the tunnelling action was ensured 
here by imposing the thermodynamic constraint of decreasing 
internal energy, alternately, the argument from stochastic inflation that
the horizon area not increase. For rotating black holes, the action of the instanton,
related to the free energy, now contains a $\beta\Omega J$ 
term \cite{Chao:1998uj,Chao:1998hk,Wu:2004db}, dependent on the angular 
momentum and a potentially arbitrary periodicity of Euclidean time. 
Further, the scalar field in the Kerr-de Sitter background will now have 
superradiant modes \cite{Press:1972zz,Teukolsky:1974yv,Tachizawa:1992ue},
that will likely have a stronger effect on the system than
any putative tunnelling decay process. We plan to study this system further.
 
The aim of this paper has been to push the theory of vacuum decay to
its limits, and yet we find nothing unreasonable in the results. It would be
of interest to use the phenomena discussed here to test the scope of theories
of quantum gravity. As for actual applications to our universe, 
vacuum decay during inflation can take place when there is a secondary,
`spectator' field present with a suitable false vacuum state. Since `flat' 
potentials are a common feature of most inflationary models of the early 
universe, the Hawking-Moss, or up-tunnelling seems the most likely
type of transition in this situation. 

\acknowledgments

This work was supported in part by the Leverhulme Trust 
[Grant No. RPG-2016-233] (RG/IGM/SP), by the
STFC [Consolidated Grant ST/P000371/1] (RG/IGM), 
by JSPS Overseas Research Fellowships (NO) and by the Perimeter Institute for 
Theoretical Physics (RG/NO). Research at Perimeter Institute is supported by 
the Government of Canada through the Department of Innovation, 
Science and Economic Development Canada and by the Province of 
Ontario through the Ministry of Research, Innovation and Science.


\begin{thebibliography}{99}

\bibitem{Hawking:1981fz} 
S.~W.~Hawking and I.~G.~Moss,
{\it Supercooled Phase Transitions in the Very Early Universe},
Phys.\ Lett.\  {\bf 110B}, 35 (1982)
[Adv.\ Ser.\ Astrophys.\ Cosmol.\  {\bf 3}, 154 (1987)].

\bibitem{CDL}
S.~Coleman and F.~De~Luccia, 
{\it {Gravitational effects on and of vacuum decay}},  
Phys.Rev. {\bf D21} (1980) 3305--3315.

%\cite{Gregory:2020cvy}
\bibitem{Gregory:2020cvy}
R.~Gregory, I.~G.~Moss and N.~Oshita,
{\it Black Holes, Oscillating Instantons, and the Hawking-Moss transition},
[\href{http://xxx.lanl.gov/abs/2003.04927}{{\tt arXiv:2003.04927 [hep-th]}}].

\bibitem{coleman1977}
S.~Coleman, 
{\it {Fate of the false vacuum: Semiclassical theory}},  
Phys.Rev. {\bf D15} (1977) 2929--2936.

\bibitem{callan1977}
C. G. Callan and S.~Coleman,
{\it {Fate of the false vacuum II: First quantum corrections}},  
Phys.Rev. {\bf D16} (1977) 1762--1768.

\bibitem{Vilenkin:1982de}
A.~Vilenkin,
{\it Creation of Universes from Nothing},
Phys. Lett. B \textbf{117}, 25-28 (1982)

\bibitem{Vilenkin:1983xq}
A.~Vilenkin,
{\it The Birth of Inflationary Universes},
Phys. Rev. D \textbf{27}, 2848 (1983)

\bibitem{Hartle:1983ai}
J.~B.~Hartle and S.~W.~Hawking,
{\it Wave Function of the Universe},
Phys.\ Rev.\ D {\bf 28}, 2960 (1983);
Adv. Ser. Astrophys. Cosmol. \textbf{3}, 174-189 (1987)

\bibitem{Starobinsky:1986fx}
A.~A.~Starobinsky,
{\it Stochastic De Sitter (inflationary) Stage In The Early Universe}
Lect.\ Notes Phys.\  {\bf 246} (1986) 107.

\bibitem{Starobinsky:1994bd}
A.~A.~Starobinsky and J.~Yokoyama,
{\it Equilibrium state of a selfinteracting scalar field in the De Sitter background},
Phys. Rev. D \textbf{50}, 6357-6368 (1994)
[\href{http://xxx.lanl.gov/abs/astro-ph/9407016}{{\tt  astro-ph/9407016}}].

\bibitem{Linde:1993nz}
A.~D.~Linde and A.~Mezhlumian,
{\it Stationary universe},
Phys. Lett. B \textbf{307}, 25-33 (1993)
[\href{http://xxx.lanl.gov/abs/gr-qc/9304015}{{\tt  gr-qc/9304015}}].

\bibitem{Li:2007uc}
M.~Li and Y.~Wang,
{\it A Stochastic Measure for Eternal Inflation},
JCAP \textbf{08}, 007 (2007)
[\href{http://xxx.lanl.gov/abs/0706.1691}{{\tt arXiv:0706.1691 [hep-th]}}].

\bibitem{Brown:2007sd}
A.~R. Brown and E.~J. Weinberg, 
{\it {Thermal derivation of the Coleman-De Luccia tunneling prescription}},
Phys.Rev. {\bf D76} (2007) 064003,
[\href{http://xxx.lanl.gov/abs/0706.1573}{{\tt arXiv:0706.1573 [hep-th]}}].

\bibitem{Gregory:2013hja}
R.~Gregory, I.~G.~Moss and B.~Withers,
{\it {Black holes as bubble nucleation sites}},  
JHEP {\bf 1403} (2014) 081,
[\href{http://xxx.lanl.gov/abs/1401.0017}{{\tt arXiv:1401.0017 [hep-th]}}].

\bibitem{Burda:2015isa} 
P.~Burda, R.~Gregory and I.~Moss,
{\it Gravity and the stability of the Higgs vacuum},
Phys.\ Rev.\ Lett.\  {\bf 115}, 071303 (2015)
[\href{http://xxx.lanl.gov/abs/1501.04937}{{\tt arXiv:1501.024937 [hep-th]}}].

\bibitem{Burda:2015yfa}
P.~Burda, R.~Gregory and I.~Moss,
{\it Vacuum metastability with black holes},
JHEP {\bf 1508}, 114 (2015)
[\href{http://xxx.lanl.gov/abs/1503.07331}{{\tt arXiv:1503.07331 [hep-th]}}].

\bibitem{Burda:2016mou} 
P.~Burda, R.~Gregory and I.~Moss,
{\it The fate of the Higgs vacuum},
JHEP {\bf 1606}, 025 (2016)
[\href{http://xxx.lanl.gov/abs/1601.02152}{{\tt arXiv:1601.02152 [hep-th]}}].

\bibitem{Chadburn:2013mta} 
S.~Chadburn and R.~Gregory,
{\it Time dependent black holes and scalar hair},
Class.\ Quant.\ Grav.\  {\bf 31}, no. 19, 195006 (2014)
[\href{http://xxx.lanl.gov/abs/1304.6287}{{\tt arXiv:1304.6287 [gr-qc]}}].

\bibitem{Gregory:2017sor} 
R.~Gregory, D.~Kastor and J.~Traschen,
{\it Black Hole Thermodynamics with Dynamical Lambda},
JHEP {\bf 1710}, no. 10, 118 (2017)
[\href{http://arxiv.org/abs/1707.06586}{{\tt arXiv:1707.06586 [hep-th]}}.

%\cite{Gregory:2018ghc}
\bibitem{Gregory:2018ghc}
R.~Gregory, D.~Kastor and J.~Traschen,
{\it Evolving Black Holes in Inflation},
Class. Quant. Grav. \textbf{35}, no.15, 155008 (2018)
[\href{http://arxiv.org/abs/1804.03462}{{\tt arXiv:1804.03462 [hep-th]}}.

\bibitem{Kastor:2009wy} 
D.~Kastor, S.~Ray and J.~Traschen,
{\it Enthalpy and the Mechanics of AdS Black Holes},
Class.\ Quant.\ Grav.\  {\bf 26}, 195011 (2009)
[\href{http://xxx.lanl.gov/abs/0904.2765}{{\tt arXiv:0904.2765 [hep-th]}}].

\bibitem{Dolan:2012jh} 
B.~P.~Dolan,
{\it Where Is the PdV in the First Law of Black Hole Thermodynamics?},
[\href{http://xxx.lanl.gov/abs/1209.1272}{{\tt arXiv:1209.1272 [gr-qc]}}].

\bibitem{Dolan:2013ft} 
B.~P.~Dolan, D.~Kastor, D.~Kubiz\v n\'ak, R.~B.~Mann and J.~Traschen,
{\it Thermodynamic Volumes and Isoperimetric Inequalities for de Sitter Black Holes},
Phys.\ Rev.\ D {\bf 87}, no. 10, 104017 (2013)
[\href{http://xxx.lanl.gov/abs/1301.5926}{{\tt arXiv:1301.5926 [hep-th]}}].

\bibitem{Kubiznak:2016qmn}
D.~Kubiz\v n\'ak, R.~B.~Mann and M.~Teo,
{\it Black hole chemistry: thermodynamics with Lambda},
Class.\ Quant.\ Grav.\  {\bf 34}, no. 6, 063001 (2017)
[\href{http://xxx.lanl.gov/abs/1608.06147}{{\tt arXiv:1608.06147 [hep-th]}}].

\bibitem{Gregory:2019dtq}
R.~Gregory and A.~Scoins,
{\it Accelerating Black Hole Chemistry},
Phys. Lett. B \textbf{796}, 191-195 (2019)
[\href{http://xxx.lanl.gov/abs/1904.09660}{{\tt arXiv:1904.09660 [hep-th]}}].

\bibitem{Linde:1991sk}
A.~D.~Linde,
{\it Hard art of the universe creation (stochastic approach to tunneling 
and baby universe formation)},
Nucl. Phys. B \textbf{372}, 421-442 (1992)
[\href{http://xxx.lanl.gov/abs/hep-th/9110037}{{\tt  hep-th/9110037}}].

\bibitem{Liddle:1993fq} 
A.~R.~Liddle and D.~H.~Lyth,
{\it The Cold dark matter density perturbation},
Phys.\ Rept.\  {\bf 231}, 1 (1993)
[\href{http://xxx.lanl.gov/abs/astro-ph/9303019}{{\tt  astro-ph/9303019}}].

\bibitem{Liddle:1994dx} 
A.~R.~Liddle, P.~Parsons and J.~D.~Barrow,
{\it Formalizing the slow roll approximation in inflation}
Phys.\ Rev.\ D {\bf 50}, 7222 (1994)
[\href{http://xxx.lanl.gov/abs/astro-ph/9408015}{{\tt  astro-ph/9408015}}].

\bibitem{Oshita:2017hsb}
N.~Oshita,
{\it Generalized second law of thermodynamics and cosmological decoherence},
Phys. Rev. D \textbf{97}, no.2, 023510 (2018)
[\href{http://xxx.lanl.gov/abs/1709.08807}{{\tt arXiv:1709.08807 [gr-qc]}}].

\bibitem{Oshita:2019jan} 
N.~Oshita, K.~Ueda and M.~Yamaguchi,
{\it Vacuum decays around spinning black holes},
JHEP {\bf 2001}, 015 (2020)
[\href{http://xxx.lanl.gov/abs/1909.01378}{{\tt arXiv:1909.01378 [hep-th]}}].

\bibitem{Chao:1998uj}
W.~Z.~Chao,
{\it Quantum creation of a black hole},
Int. J. Mod. Phys. D \textbf{6} (1997), 199
[\href{https://arxiv.org/abs/gr-qc/9801020v1}{arXiv:gr-qc/9801020 [gr-qc]}].

\bibitem{Chao:1998hk}
W.~Z.~Chao,
{\it Pair creation of black holes in Anti-de Sitter space background. 2},
Phys. Lett. B \textbf{445} (1999), 274
[\href{https://arxiv.org/abs/gr-qc/9810012}{arXiv:gr-qc/9810012 [gr-qc]}].

\bibitem{Wu:2004db}
Z.~C.~Wu,
{\it Creation of Kerr-de Sitter black hole in all dimensions},
Phys. Lett. B \textbf{613} (2005), 1-4
[\href{https://arxiv.org/abs/gr-qc/0412041}{arXiv:gr-qc/0412041 [gr-qc]}].

\bibitem{Press:1972zz}
W.~H.~Press and S.~A.~Teukolsky,
{\it Floating Orbits, Superradiant Scattering and the Black-hole Bomb},
Nature \textbf{238}, 211-212 (1972)

\bibitem{Teukolsky:1974yv}
S.~A.~Teukolsky and W.~H.~Press,
{\it Perturbations of a rotating black hole. III - Interaction of the hole with 
gravitational and electromagnetic radiation},
Astrophys. J. \textbf{193}, 443-461 (1974)

\bibitem{Tachizawa:1992ue}
T.~Tachizawa and K.~i.~Maeda,
{\it Superradiance in the Kerr-de Sitter space-time},
Phys. Lett. A \textbf{172}, 325-330 (1993)


\end{thebibliography}
\end{document}